\def\Roman#1{\uppercase\expandafter{\romannumeral#1}}
\documentclass[a4paper,12pt]{article}
\usepackage{cite}
\usepackage{amssymb,amsfonts}
\usepackage{amsmath}
\usepackage[mathscr]{eucal}
\usepackage{mathrsfs}
\usepackage[utf8x]{inputenc}

\title{Path integrals on a manifold that is a product of the total space of the principal fiber bundle and the vector space}

\author{S. N. Storchak\\
\small{ A. A. Logunov Institute for High Energy Physics}\\
\small{of NRC ``Kurchatov Institute'',}\\
\small{Protvino, 142281, Russian Federation,}\\
}

\begin{document}

  \maketitle

\begin{abstract}
Using  the path integral measure factorization method  based on the nonlinear filtering equation from the stochastic process theory, we consider the reduction procedure in  Wiener path integrals for a mechanical system with symmetry that describes the motion of two interacting scalar particles  on a special smooth compact Riemannian manifold - the product of total space of the principal fiber bundle and the vector space.  The original manifold, the configuration space of this system, is endowed with an isometric free proper action of a compact semisimple unimodular Lie group. The proposed reduction procedure leads to the integral relation between path integrals that  represent fundamental solutions of the inverse Kolmogorov equations on the initial and reduced manifolds.
For the case of reduction onto the zero-momentum level, the reduction Jacobian is obtained, which is an additional potential term to the Hamiltonian.
\end{abstract}

\section{Introduction}
In this paper we consider the path integral ``quantization'' of a special mechanical system with symmetry. This system describes the motion of two interacting scalar particles on the product manifold consisting of a smooth compact finite-dimensional Riemannian manifold and a finite-dimensional vector space. In the studied problem, a free isometric smooth action of a compact semisimple unimodular Lie group is given on the original manifold. We also require that the system be invariant with respect to the actions of this group.

From the symmetry of our problem it follows that configuration space of the mechanical system (the original manifold) can be regarded as the total space of the principal fiber bundle $\pi':\mathcal P\times \mathcal V\to \mathcal P\times_{\mathcal G} \mathcal V.$ This means that the configuration space of the reduced mechanical system is $\mathcal P\times_{\mathcal G} \mathcal V$, which  is the total space of the associated bundle for the principal fiber bundle $\rm P(\mathcal M,\mathcal G)$.

The main problem we will study in the paper is the behaviour of the original path integral under the reduction performed in our system,
namely, the transformation of the path integral when the initial space is changed for the reduced one.

Earlier in our works \cite{Storchak_1,Storchak_11,Storchak_12,Storchak_2}, this problem was investigated for a simpler case - the motion of a scalar particle  on a manifold with a given group action. There, the Wiener path integrals representing solutions of the backward Kolmogorov equations were considered.
These path integrals were determined by the method proposed  by  Belopolskaya and Dalecky in \cite{Dalecky_1,Dalecky_2}. In accordance with this method, the measure of the path integral is generated by a stochastic process, which is a solution to the stochastic differential equation given on the manifold. 

The main advantage of this method is the peculiar use and determination of local stochastic differential equations   on charts of the manifold.  Local stochastic differential equations are defined using exponential mappings and therefore are the Stratonovich equations. This is necessary for transforming local equations in a covariant way in case of  changing the charts.
On overlapping of the charts the local stochastic differential equations and their solutions transform  into each other.
This property enables one  to define a global stochastic process from the  local processes and therefore, the path integral on a manifold representing a  global evolution semigroup. Local evolution semigroups, the superposition of which leads to a global semigroup, are represented by path integrals over measures defined by local stochastic differential equations.
This allows us to deduce the transformation of the path integral on a manifold from the transformation of path integrals for local evolutionary semigroups. In turn, these transformation are actually based on the transformations of local stochastic differential equations.  Note that the similar method is also valid for determining  stochastic processes on the vector and principal bundle \cite{Dalecky_2}.

Belopolskaya and Dalecky's approach to the definition of stochastic processes (and the corresponding semigroups) was used in our works, where we studied the procedure of reduction of path integrals. There it was discovered \cite{Storchak_1,Storchak_11} that the reduction of the path integrals for a mechanical systems with symmetry, which leads to the factorization of the path integral measure, can be performed using the nonlinear filtering equation from the theory of the stochastic  processes.
A similar approach was later developed in \cite{Elworthy}.

The main result of our previous work on the reduction of path integrals is the conclusion about the non-invariance of the path measure during reduction, the calculation of the corresponding reduction Jacobian and its geometric representation. 

The mechanical system that we used in our studies of the reduction procedure in path integrals is  similar by their geometric properties to the pure Yang-Mills  theory. The initial configuration spaces in both dynamic systems are the principal fiber bundles.
In addition, the reduction leads to the orbit space where reduced evolution takes place. Therefore, with regard to the path integral quantization, the mechanical system can, in a sense, be considered  as a finite-dimensional model of the Yang-Mills theory.

The choice of the mechanical system to study  the path integral reduction procedure in the present article is due to the fact that this system is an analog of a field system that describes the interaction of a Yang-Mills field with a scalar field.
Note that in the case of the interaction of the electromagnetic field with the scalar field, a similar geometric representation  was in used \cite{Huffel-Kelnhofer}

\section{Definitions}
We  consider the path integrals representing the solution of the backward Kolmogorov equation given on a smooth compact Riemannian manifold $ \tilde {\mathcal P} = \mathcal P \times \mathcal V $:
\begin{equation}
\left\{
\begin{array}{l}
\displaystyle
\left(
\frac \partial {\partial t_a}+\frac 12\mu ^2\kappa \bigl[\triangle
_{\cal P}(p_a)+\triangle
_{\cal V}(v_a)\bigr]+\frac
1{\mu ^2\kappa m}V(p_a,v_a)\right){\psi}_{t_b} (p_a,t_a)=0,\\
{\psi}_{t_b} (p_b,v_b,t_b)=\phi _0(p_b,v_b),
\qquad\qquad\qquad\qquad\qquad (t_{b}>t_{a}),
\end{array}\right.
\label{1}
\end{equation}
where $\mu ^2=\frac \hbar m$ , $\kappa $  is a real positive
parameter,  $V(p,f)$ is the group-invariant potential term:
 $V(pg,g^{-1}v)=V(p,v)$, $g\in \mathcal G$, 
$\triangle _{\cal P}(p_a)$ is the Laplace--Beltrami operator on a
manifold $\cal P$ and $\triangle
_{\cal V}(v)$ is the Laplacian on the vector space $\cal V$. In the local chart $(U_{\cal P}\times U_{\cal V},\varphi )$, $\varphi=(\varphi^A,\varphi^a)$, of the manifold $\mathcal P$, where the point $(p,v)$ is represented by the  coordinates $(Q^A,f^a)$, $\triangle _{\cal P}$ has the following form\footnote{In  our formulas  we  assume that there is 
sum over the repeated indices. The indices denoted by the capital
letters  ranged from 1 to $n_{\cal P}=\rm{dim} \cal P$, and the small Latin letters, except $i,j,k,l$, -- from 1 to $n^{\cal V}=\dim \cal V$.}
\begin{equation}
\triangle _{\cal P}(Q)=G^{-1/2}(Q)\frac \partial {\partial
Q^A}G^{AB}(Q)
G^{1/2}(Q)\frac\partial {\partial Q^B},
\label{2}
\end{equation}
where $G^{AB}(Q)$ are the components of the matrix which is inverse
to the matrix $G_{AB}$ of the components of the initial Riemannian
metric given  in the coordinate basis $\{\frac{\partial}{\partial
Q^A}\}$, $G=det (G_{AB})$.

In the same chart, $\triangle_{\cal V}$ is given by 
\[
 \triangle_{\cal V}(f)=G^{ab}\frac{\partial}{\partial f^a\partial f^b},
\]
where matrix $G^{ab}$ is inverse to the matrix $G_{ab}$ representing the metric on $\mathcal V$.
By the assumption used in the paper, the matrix $G_{ab}$ consists of fixed constant elements. It is admitted that $ G_{ab}$ may have off-diagonal elements.

We also assume that the coefficients and the initial function of
equation (\ref{1}) are properly bounded and satisfy the necessary smooth
requirement, so that the solution of  equation, as it follows from
\cite{Dalecky_1}, can be represented as follows:
\begin{eqnarray}
{\psi}_{t_b} (p_a,v_a,t_a)&=&{\rm E}\Bigl[\phi _0(\eta_1 (t_b),\eta_2(t_b))\exp \{\frac
1{\mu
^2\kappa m}\int_{t_a}^{t_b}V(\eta_1(u),\eta_2(u))du\}\Bigr]\nonumber\\
&=&\int_{\Omega _{-}}d\mu ^\eta (\omega )\phi _0(\eta (t_b))\exp
\{\ldots 
\},
\label{orig_path_int}
\end{eqnarray}
where ${\eta}(t)$ is a global stochastic process on a manifold 
$\tilde{\cal P}=\cal P\times \cal V$, formed from the processes $\eta_1(t)$ and $\eta_2(t)$; ${\mu}^{\eta}$ is the path integral measure  on
the path space $\Omega _{-}=\{\omega (t)=\omega^1(t)\times\omega^2(t):\omega^{1,2} (t_a)=0, \eta_1
(t)=p_a+\omega^1 (t),\eta_2(t)=v_a+\omega^2(t)\}$ given on manifold $\tilde {\mathcal P}$. This meaure is defined by the probability distribution of a stochastic  process $\eta(t)$.

In a local chart $(U_{\cal P}\times U_{\cal V},\varphi )$ of the manifold $\tilde {\mathcal P}$, the process $\eta (t)$ is given by the solution of two stochasic differential equations:
\begin{equation}
d\eta_1^A(t)=\frac12\mu ^2\kappa G^{-1/2}\frac \partial {\partial
Q^B}(G^{1/2}G^{AB})dt+\mu \sqrt{\kappa }{\cal X}_{\bar{M}}^A(\eta_1
(t))dw^{
\bar{M}}(t),\\
\label{eta_1}
\end{equation}
and
\begin{equation}
 d\eta_2^a(t)=\mu \sqrt{\kappa }{\cal X}_{\bar{a}}^b 
dw^{
\bar{b}}(t)\\
\label{eta_2}
\end{equation}
(${\cal X}_{\bar{M}}^A$ and ${\cal X}_{\bar{a}}^b$ are  defined  by the local equalities
$\sum^{n_{\mathcal P}}_{\bar{\scriptscriptstyle K}\scriptscriptstyle =1}{\cal
X}_{\bar{K}}^A{\cal X}_{\bar{K}}^B=G^{AB}$ and  $\sum^{n_{\mathcal V}}_{\bar{\scriptscriptstyle a}\scriptscriptstyle =1}{\cal
X}_{\bar{a}}^b{\cal X}_{\bar{a}}^c=G^{bc}$, $dw^{\bar{M}}(t)$ and $dw^{\bar{b}}(t)$ are the independent Wiener processes.
Here and what follows  we  denote the Euclidean
indices by over-barred indices). Note that equations (\ref{eta_1}) and (\ref{eta_2}) are the Stratonovich equations.

We will assume that  equation (\ref{1}) has a fundamental solution -- the Green function 
$G_{\tilde{\cal P}}(p_b,v_b,t_b;p_a,v_a,t_a)$ , which is the kernel of the semigroup 
\[
\psi (p_a,v_a,t_a)=\int G_{\tilde{\cal P}}(p_b,v_b,t_b;p_a,v_a,t_a)\phi _0(p_b,v_b)dv_{\tilde{\cal P}}(p_b,v_b),
\]
where $dv_{\tilde{\cal P}}(p,v)$ is a volume element on the manifold $\tilde{\cal P}$.

To get a probabilistic representation of the kernel $ G_{\tilde{\cal P}},$ of the semigroup it is necessary to replace $\varphi_0 $ with the delta function in (\ref{orig_path_int}).
The same can be done if we consider
the  limit of the corresponding functions that approximate the delta function.

The choice of the coefficients of equation (\ref{1}) from a certain class of functions allows us to use the fundamental solution as a solution to the forward Kolmogorov equation. Also note that the  Schr\"{o}dinger equation can be obtained from  the forward Kolmogorov equation if  we replace $ \kappa $ for $i$. But this does not mean that Feynman integrals can be obtained using this method from Wiener-type integrals. This transition needs special research.

 The global semigroup determined by equation (\ref{orig_path_int})  is defined in \cite{Dalecky_1, Dalecky_2} by the limit (under the refinement of the time interval) of the superposition of the local semigroups. In our case it is given by
\begin{equation}
\!\psi _{t_b}(p_a,v_a,t_a)=U(t_b,t_a)\phi _0(p_a,v_a)=
{\lim}_q {\tilde U}_{\eta}(t_a,t_1)\cdot\ldots\cdot
{\tilde U}_{\eta}(t_{n-1},t_b)
\phi _0(p_a,v_a),
\label{6}
\end{equation}
where  each of the local semigroup  
${\tilde U}_{\eta}$ is as follows:
\[
 {\tilde U}_{\eta}(s,t) \phi (p,v)={\rm E}_{s,p,v}\phi (\eta_1
 (t),\eta_2(t))\,\,\,\,\,\,
 s\leq t\,\,\,\,\,\,\eta_1 (s)=p,\;\eta_2 (s)=v.
 \]
These local semigroups are also given by the path integrals with the integration measures determined by the local representatives
 $\varphi
 ^{\tilde{\cal P}}(\eta_t)
\equiv\{\eta ^A_1 (t),\eta^a_2(t)\}$ of the global stochastic process $\eta(t)$.

In the following,
we derive the transformation properties of the path integral  (\ref{orig_path_int}) by studying  the transformation of the local semigroups ${\tilde U}_{\eta}$. 
 Since  the potential term of the Hamiltonian operator  inessential when performing path integral transforms, it will be omitted during this process. We will  recover it in our final formulae.

\section{Principal fiber bundle coordinates}
On the Riemannian manifold $\tilde{\mathcal P}=\mathcal P \times \mathcal V$, the configuration space of our initial mechanical system, we are given a smooth isometric free and proper action of a compact semisimple Lie group $\mathcal G$: $(\tilde p,\tilde v)=(p,v)g=(pg,g^{-1}v)$. That is, we were given a right action of the group $\mathcal G$ on $\tilde{\mathcal P}$. In a local coordinates $(Q^A,f^a)$, this action is given as follows:
\[
 {\tilde Q}^A=F^A(Q,g),\;\;\;\;{\tilde f}^b=\bar D^b_a(g)f^a,
\]
where $\bar D^b_a(g)\equiv D^b_a(g^{-1})$,
and by $D^b_a(g)$ we denote the matrix of  the finite-dimensional representation of the group $\mathcal G$
acting on the vector space $\mathcal V$. In particularly,

In coordinates, the right action of the group $\mathcal G$ on $\mathcal P$ means that
\[
 F(F(Q,g_1),g_2)=F(Q,\rm \hat {\Phi}(g_1,g_2)),
\]
where the function $\rm\hat {\Phi}$  determines the group multiplication law in the space of the group parameters.

 The Riemannian metric of the manifold $\tilde{\mathcal P}$ can be written as follows:
\begin{equation}
 ds^2=G_{AB}(Q)dQ^AdQ^B+G_{ab}\,df^adf^b.
\label{metr_orig}
\end{equation}
We note, that the components of the metric are not arbitrary, but must satisfy certain relations due to the isometric action of the group $\mathcal G$ on $\tilde{\mathcal P}$.
 
In  particularly, the metric tensor $G_{AB}(Q)$ satisfies the following relation:
\begin{equation}
 G_{AB}(Q)=G_{DC}(F(Q,g))F^D_A(Q,g)F^C_B(Q,g),
\label{relat_G_AB}
\end{equation}
with $ F^B_A(Q,g)\equiv \partial F^B(Q,g)/\partial Q^A$.
A similar relation for the tensor $G_{pq}$,
\begin{equation}
 G_{pq}=G_{ab}\bar D^a_p(g)\bar D^b_q(g),
\label{relat_g_ab}
\end{equation}
can be derived from the linear isometrical action of the group $\mathcal G$ in the vector space $\mathcal V$.

The Killing vector fields for the  metric (\ref{metr_orig}) given on the  manifold $\tilde{\mathcal P}$ are   the  vector fields on  $\mathcal P$ and $\mathcal V$.  In local coordinates $(Q^A,f^b)$,  they are    representedas as
 $K^A_{\alpha}(Q)\partial / \partial Q^A$
 with
$K^A_{\alpha}(Q)=\partial {\tilde Q}^A/\partial a^{\alpha}|_{a=e}$ and
$K^b_{\alpha}(f)\partial/\partial f^b$ with 
$$K^b_{\alpha}(f)=\partial {\tilde f}^b/\partial a^{\alpha}|_{a=e}=\partial {\bar D}^b_c(a)/\partial a^{\alpha}|_{a=e}f^c\equiv({\bar J}_{\alpha})^b_c f^c.$$  
The generators ${\bar J}_{\alpha}$ of the representation ${\bar D}^b_c(a)$ satisfy the  commutation relation 
$[{\bar J}_{\alpha},{\bar J}_{\beta}]={\bar c}^{\gamma}_{\alpha \beta}{\bar J}_{\gamma}$, where the structure constants
${\bar c}^{\gamma}_{\alpha \beta}=-{c}^{\gamma}_{\alpha \beta}$.


Based on \cite{AbrMarsd}, we conclude that the  action of a group $\mathcal G$ on the manifold  $\tilde{\mathcal P}$  leads to the  principal fiber bundle $ \pi':\mathcal P\times \mathcal V \to \mathcal P\times _{\mathcal G}\mathcal V$.\footnote{$\pi':(p,v)\to [p,v]$, where  $[p,v]$ is the equivalence class formed  by the  relation   $(p,v)\sim (pg,g^{-1}v).$} This allow us to regard the initial manifold $\tilde{\mathcal P}$ as a total space of this principal fiber bundle $\rm P(\tilde{\mathcal M},\mathcal G)$. By $\tilde{\mathcal M}$ we denoted the orbit space manifold $\mathcal P\times _{\mathcal G}\mathcal V$ -- the base space of the bundle $ \pi'$.
From this it follows that we can express the local coordinates $(Q^A, f^n)$ of the point $(p,v)\in \tilde{\mathcal P}$ in terms of the  coordinates of the principal fiber bundle.

As coordinates in this  principal bundle we will  use the adapted coordinates. 
Such coordinates for the principal bundle were used, for example, in \cite{Creutz, Razumov_1, Razumov_2, Razumov_3, Plyush, Storchak_11, Storchak_12,Storchak_2, Huffel-Kelnhofer}.                          
As for the coordinates for the principal  bundle similar to that given in this article, we can  refer to \cite{Huffel-Kelnhofer,Storchak_11,Storchak_12,Storchak_4,Storchak_5,Storchak_6}. Note that the main application  of adapted coordinates are the calculations in gauge field theories.

Adapted coordinates consist of coordinates associated with the base space of the principal bundle and the remaining coordinates dealing with the group manifold. Also in their definition, the main role is played by the local submanifold of the total space of the bundle. This submanifold  has  the transversal intersection with each of the orbits.
When determining the adapted coordinates for our principal fiber bundle $\rm P(\tilde {\mathcal M}, \mathcal G)$, such local surface $\Sigma$ is chosen in the totall space of the principal fiber bundle $ \rm P(\mathcal M, \mathcal G)$.

In this case, the local section $\sigma_i$ of $ \rm P(\mathcal M, \mathcal G)$, $\sigma_i:U_i\to\Sigma_i \in\pi_{\rm P}^{-1}(U_i)$, $x=\pi_{\rm P}(p)$ are used to define the local section $\tilde \sigma_i$ of $\rm P(\tilde {\mathcal M}, \mathcal G)$, $\pi' \cdot\tilde \sigma_i = \rm{id}$:  
\[
 \tilde \sigma_i([p,v])=(\sigma_i(x),a(p) v), 
\]
where  $a(p)$ is the group element such that $p=\sigma_i(x)a(p)$.
Thus,  $\tilde \sigma_i$  sends $[p,v]$ to some element $(\tilde p,\tilde v)\in \mathcal P\times V$. 

From
\[
 (\sigma_i(x),a(p)\, v)=(p\,a^{-1}(p),a(p)\, v)=(p,v)\,a^{-1}(p),
\]
it follows that
\[
 \tilde \sigma_i([p,v])=(p,v)\,a^{-1}(p).
\]
This is similarly to the case of the principal fiber bundle $ \rm P(\mathcal M, \mathcal G)$.
Therefore we can say 
 that the image of $\tilde \sigma_i$ is a local surface $\tilde \Sigma_i$ in $\mathcal P\times \mathcal V$.

Due to 
 the local isomorphisms of the principal fiber bundle ${\rm P}(\mathcal P\times _{\mathcal G}V,\mathcal G)$ and the  trivial principal bundles $\tilde \Sigma_i\times \mathcal G \to \tilde \Sigma_i$ we can
 introduce a new atlas on ${\rm P}(\mathcal P\times _{\mathcal G}V,\mathcal G)$  with charts that are related to the submanifolds $\{\tilde \Sigma_i\}$.
The coordinate functions  of these charts $(\tilde U_i,\tilde\varphi_i)$, where $\tilde U_i$ is an open neighborhood of the point $[p,v]$ given on the base space $\mathcal P\times_{\mathcal G}V$, are such that
\[
 \tilde \varphi_i^{-1} : \pi'^{-1}(\tilde U_i)\to\tilde \Sigma_i \times \mathcal G\;\; {\rm  and}
\]
\[
 \!\!\!\!\!\!\!\tilde \varphi_i:\tilde \Sigma_i\times \mathcal G\to \pi'^{-1}(\tilde U_i).
\]
To implement this general scheme for determining adapted coordinates, we must first present in coordinates its main elements. The local submanifold $\Sigma $ in $\mathcal P$ is given by the system of equations
$\chi^{\alpha}(Q)=0,\,\alpha=1,...,n^{\mathcal G}$. So, the point with the coordinates $Q^{\ast A}$, $\{ \chi^{\alpha}(Q^{\ast A})=0\}$,   belongs to  $\Sigma $. Note that the coordinates $Q^{\ast A}$ are dependent coordinates.

The group coordinates $a^{\alpha}(Q)$ of a point $p\in \mathcal P$ are defined by the solution of the following equation:
\[
 \chi^{\alpha}(F^A(Q,a^{-1}(Q)))=0.
\]
This group element  carries the point $p$ to the submanifold  $\Sigma$, so that
\[
 Q^{\ast A}=F^A(Q,a^{-1}(Q))).
\]
 
The coordinates of a point $(p,v)\in \tilde{\Sigma}$ are $(Q^{\ast A},\tilde f^a)$.
Therefore, the coordinate functions look as follows:

\[
 \tilde \varphi_i^{-1} :(Q^A,f^b)\to (Q^{\ast}{}^A(Q),\tilde f^b(Q),a^{\alpha}(Q)\,),
\]
where
\[
\tilde f^b(Q) = D^b_c(a(Q))\,f^c,
\]
($\bar D^b_c(a^{-1})\equiv D^b_c(a))$.

The coordinate function $\tilde \varphi_i$ 
\[
 \tilde \varphi_i :(Q^{\ast}{}^B,\tilde f^b,a^{\alpha})\to (F^A(Q^{\ast},a), \bar D^c_b(a)\tilde f^b).
\]
Thus, we have demonstraited how can be defined  the special local bundle coordinates
 $(Q^{\ast}{}^A,\tilde f^b, a^{\alpha})$ 
on the principal fiber bundle 
$\pi':\mathcal P\times V\to \mathcal P\times_{\mathcal G} V$. These adapted coordinates include the dependent coordinates $Q^{\ast A}$.

Note that, in principle, independent local coordinates can also be determined in our bundle. This is the case when the local submanifold $\Sigma$ can be define  parametrically: $Q^A=Q^{\ast A}(x^i)$. The invariant coordinates $x^i$, $i=1,...,n^{\mathcal M}$, which can be identified with the coordinates given on the base manifold $\mathcal M$, together with the group coordinates $a^{\alpha}$ are used in this case as  coordinates of the point $p\in \mathcal P$. It is also necessary that the equality $n^{\mathcal M}+n^{\mathcal G}=n^{\mathcal P}$ holds true. If so, then the invariant coordinates  $x^i(Q)$ are determined from the equation
 \[
 Q^{\ast A}(x^i)=F^A(Q,a^{-1}(Q))).
\]
In addition, now we have the following equality: $\chi^{\alpha}(Q^{\ast A}(x^i))=0$. So,  the point $(p,v)\in \tilde{\mathcal P}$, whose  coordinates were $(Q^A,f^b)$, obtains the adapted coordinates  $(x^i(Q),\tilde f^a(f), a^{\alpha}(Q))$. The replacement of the coordinates $(Q^A,f^b)$ of a point $(p,v)$ for a new coordinates is performed as follows:
\begin{equation}
Q^A=F^A(Q^{\ast}(x^i),a^{\alpha}),\;\;\;f^b=\bar D^b_c(a)\tilde f^c,
\label{transf_coord}
\end{equation}
where $a^{\alpha}=a^{\alpha}(Q)$ is obtained as before.
It is this adapted coordinates will be used in article.

In the sequel we will deal, in fact, only  with the local expressions that are  given on a  separate chart. 
We assume that these local expressions can be "glued" and  we are able to restore the global expressions.

It is not difficult to obtain the representation for the Riemannian metric  given on $\mathcal P\times \mathcal V$ in terms of the principal bundle  coordinates $(x^i,{\tilde f}^b, a^{\alpha})$ which we have just introduced on the principal fiber bundle. This can be made by taking the following differentials:
\begin{eqnarray*}
 && dQ^a=F^A_BQ^{\ast}{}^B_jdx^j+F^A_{\alpha}da^{\alpha}
\nonumber\\
&&df^c=\frac{\partial \bar D^c_b(a)}{\partial a^{\alpha}}\tilde f^b da^{\alpha}+\bar D^c_b(a)d\tilde f^b,
\end{eqnarray*}
where 
 $F^A_{\alpha}(Q,a)\equiv \frac{\partial F^A}{\partial
a^{\alpha}}(Q,a)$,
$Q^{*B}_i(x)\equiv \frac{\partial Q^{*B}(x)}{\partial x^i}$.

Because of the equality
\[
F_{\alpha }^A(Q,g) = \bar u_\alpha ^\beta (g) F_{B}^A(Q,g) K_\beta
^B(Q),  
\]
where $\bar{u}^{\alpha}_{\beta}$ is an inverse matrix to the
matrix 
$\bar{v}^{\alpha}_{\beta}(a)=\frac{\partial {\Phi}^{\alpha}(b,a)}
{\partial b^{\beta}}\bigl|_{b=e}$ ($\Phi$ is a group function which
defines the group multiplication in the space of group parameters),
the  differential $df^c$ can be rewritten as
$$df^c=K^r_{\beta}(\tilde f) \bar D^c_r(a)\bar u^{\beta}_{\mu}da^{\mu}+\bar D^c_b(a)d\tilde f^b,$$
where $K^r_{\beta}(\tilde f) =(\bar J_{\beta})^r_b\tilde f^b$ is the component of the Killing vector field.

The replacement of the coordinates $(Q^A,f^a)$ for the adapted coordinates  $(x^i,\tilde f^b,a^{\alpha})$
 leads to the following components
of the Riemannian metric defined on the manifold $\tilde{\mathcal P}$ 
in the basis
$\{\partial/\partial x^i, \partial/\partial \tilde f^b, \partial/\partial a^{\alpha}\}:$
\begin{eqnarray*}
G_{ij}(x,a) &=& Q_{i}^{*A}(x) G_{AB}(Q^*(x)) Q_{j}^{*B}(x) = 
G_{ij}(x,e), 
\nonumber\\
G_{i\beta }(x,a) &=& Q_{i}^{*A}(x) G_{AB}(Q^*(x)) K_\delta ^B(Q^*(x))
{\bar u}_{\beta}^{\delta} (a) = G_{i\delta }(x,e) {\bar u}_{\beta}^{\delta} (a), 
\nonumber\\
G_{\alpha \beta }(x,\tilde f, a) &=& (G_{AB}K^A_{\mu}K^B_{\nu}+G_{rp}K^r_{\mu}K^p_{\nu}){\bar u}_{\alpha}^{\mu}(a)  {\bar u}_{\beta}^{\nu}(a)
\nonumber\\
&=&({\gamma}_{\mu\nu}(x))+{\gamma'}_{\mu\nu}(\tilde f)){\bar u}_{\alpha}^{\mu}(a)  {\bar u}_{\beta}^{\nu}(a)\equiv
d_{\mu \nu}(x,\tilde f) \bar u_{\alpha}^{\mu}(a) \bar u_{\beta}^{\nu}(a).
\end{eqnarray*}
These components were obtained taking into account  the isometry of the action of the group $\mathcal G$ on $\tilde{\mathcal P}$. As a result we have
\begin{equation}
\displaystyle
G_{\tilde A \tilde B}=\left(
\begin{array}{ccc}
G_{ij}(x,e), & 0 & G_{i\delta }(x,e) {\bar u}_{\alpha}^{\delta} (a) \\ 
0 & G_{ab}  & G_{a\mu}(\tilde f){\bar u}^{\mu}_{\alpha}(a)\\
G_{j \delta}(x,e){\bar u}^{\delta}_{\beta}(a) & G_{b\nu}(\tilde f){\bar u}^{\nu}_{\beta}(a) & d_{\mu \nu}{\bar u}^{\mu}_{\alpha}(a){\bar u}^{\nu}_{\beta}(a)\\
\end{array}
\right)
\label{11}
\end{equation}
 
Note that this metric can also be written in terms of the components $({\mathscr A}^{\alpha}_i, {\mathscr A}^{\alpha}_p)$ of the mechanical connection  that exists in the principal fiber bundle $\rm P(\tilde{\mathcal M},\mathcal G)$. These components are determined  as follows:
$${\mathscr A}^{\alpha}_i(x,\tilde f)=d^{\alpha \beta}K^C_{\beta}G_{DC}Q^{\star}{}^D_i, \;\;\;{\mathscr A}^{\alpha}_p(x,\tilde f)=d^{\alpha \beta}K^r_{\beta}G_{rp}.$$
In this case,   the transformed Riemannian metric can be represented in the following form: 
\begin{equation}
\displaystyle
G_{\tilde A \tilde B}=
\left(
\begin{array}{ccc}
{\tilde h}_{ij}+{\mathscr A}_i^\mu {\mathscr A}_j^\nu d_{\mu \nu } & 0 & {\mathscr A}_i^\mu
d_{\mu\nu}
\bar{u}^{\nu}_{\alpha}(a) \\ 
0 & G_{ab}  & {\mathscr A}^{\mu}_a d_{\mu\nu}{\bar u}^{\nu}_{\alpha}(a)\\
{\mathscr A}^{\mu}_jd_{\mu\nu}{\bar u}^{\nu}_{\beta}(a) & {\mathscr A}^{\mu}_b d_{\mu\nu}{\bar u}^{\nu}_{\beta}(a) & d_{\mu \nu}{\bar u}^{\mu}_{\alpha}(a){\bar u}^{\nu}_{\beta}(a)\\
\end{array}
\right),
\label{transfmetric}
\end{equation}
where 
${\tilde h}_{ij}(x,\tilde f)=Q^{\ast}{}^A_i{\tilde G}^{\rm H}_{AB}Q^{\ast}{}^B_j$, and ${\tilde G}^{\rm H}_{AB}=G_{AB}-G_{AC}K^C_{\mu}d^{\mu\nu}K^D_{\nu}G_{DB}.$
Further, we will also denote expressions that include $d^{\mu\nu}$, with a tilde mark above the character associated with that expression.

The inverse matrix $G^{\tilde A \tilde B}$ to matrix (\ref{transfmetric}) is as follows:
\begin{equation}
 \displaystyle
G^{\tilde A \tilde B}=\left(
\begin{array}{ccc}
 h^{ij} & \underset{\scriptscriptstyle{(\gamma)}}{{\mathscr A}^{\mu}_m} K^a_{\mu} h^{mj} & -h^{nj}\,\underset{\scriptscriptstyle{(\gamma)}}{{\mathscr A}^{\beta}_n} \bar v ^{\alpha}_{\beta} \\
\underset{\scriptscriptstyle{(\gamma)}}{{\mathscr A}^{\mu}_m} K^b_{\mu} h^{ni} & G^{AB}N^a_AN^b_B+G^{ab} & -G^{EC}{\Lambda}^{\beta}_E{\Lambda}^{\mu}_CK^b_{\mu}\bar v ^{\alpha}_{\beta}
\\
-h^{ki}\underset{\scriptscriptstyle{(\gamma)}}{{\mathscr A}^{\varepsilon}_k}\bar v ^{\beta}_{\varepsilon} & -G^{EC}{\Lambda}^{\varepsilon}_E{\Lambda}^{\mu}_CK^a_{\mu}\bar v ^{\beta}_{\varepsilon} & G^{BC}{\Lambda}^{\alpha'}_B{\Lambda}^{\beta'}_C\bar v ^{\alpha}_{\alpha'}v ^{\beta}_{\beta'}
\end{array}
\right).
\label{invers_metric}
\end{equation}
Here ${\Lambda}^{\beta}_E=(\Phi^{-1})^{\beta}_{\mu}{\chi}^{\mu}_E$,  $h^{ij}$ is an inverse matrix to the matrix $h_{ij}=Q^{\ast A}_i G^{\rm H}_{AB}Q^{\ast B}_j$ with $$G^{\rm H}_{AB}=G_{AB}-G_{AD}K^D_{\alpha}{\gamma}^{\alpha\beta}K^C_{\beta}G_{CB}.$$ 
Also, for the mechanical connection formed from the orbit metric $\gamma_{\mu \nu}$ we use the following notation:
$$\underset{\scriptscriptstyle{(\gamma)}}{{\mathscr A}^{\mu}_m}={\gamma}^{\mu \nu}K^A_{\nu}G_{AB}Q^{\ast B}_m.$$
By $\chi^{\alpha}_B$ we denote $\chi^{\alpha}_B=\partial \chi^{\alpha}(Q)/\partial Q^B |_{Q=Q^{\ast}(x)}$, $(\Phi)^{\alpha}_{\beta}=K^A_{\beta}\chi^{\alpha}_A$ is the Faddeev-Popov matrix, $N^b_B=-K^b_{\mu}(\Phi)^{\mu}_{\nu}{\chi}^{\nu}_B \equiv -K^b_{\mu}{\Lambda}^{\mu}_B$ is one of the components of a particular projector on a tangent space to the orbit space. (This projector was defined in \cite{Storchak_4,Storchak_5}.

The determinant of matrix (\ref{transfmetric}) is equal to
\begin{eqnarray}
 \det G_{\tilde A \tilde B}=(\det d_{\alpha\beta})\,(\det {\bar u}^{\mu}_{\nu}(a))^2 \displaystyle
\det \left(
\begin{array}{cc}
\tilde h_{ij} & \tilde G^{\rm H}_{B b}Q_{i}^{*B}\\
\tilde G^{\rm H}_{Aa}Q_{j}^{*A} & \tilde G^{\rm H}_{ba}\\
\end{array}
\right),
\label{det}
\end{eqnarray}
where $\tilde G^{\rm H}_{Aa}=-G_{AB}K^B_{\mu}d^{\mu\nu}K^b_{\nu}G_{ba}$, $\tilde G^{\rm H}_{ba}=G_{ba}-G_{bc}K^c_{\mu}d^{\mu\nu}K_{\nu}^pG_{pa}$.

Note that the  determinant on the right hand side of (\ref{det}) is the determinant of the metric  defined on the orbit space $\tilde{\mathcal M}=
\mathcal P\times_{\mathcal G}\mathcal V$ of the principal fiber bundle $\rm P(\tilde{\mathcal M},\mathcal G).$
In what follows we will denote this determinant by $H$.

Also note that in the matrix $G^{\tilde A \tilde B}$,  the upper left quadrant of the matrix 
(\ref{invers_metric}) is ​​the matrix that represents the inverse metric to the metric in the orbit space of our principal fiber bundle.

\section{Transformation of the stochastic process and the semigroup resulting after changing coordinates}
Our next task should be the determination  of a new stochastic process, which should correspond to the newly obtained  coordinates on the manifold. According to the definition of the path integral (global semigroup), which we use in the article, the path integral on the manifold is obtained as the limit of the superposition of local evolution semigroups. Therefore, we can restrict ourselves to finding local semigroups that are determined by new local stochastic processes. As such a local process, we will take exactly the one that has the following components $(x^i(t),\tilde f^a(t),a^{\alpha}(t))$. A local process with such a components is a local representative of the global process $ \zeta(t) $ on the manifold $ \tilde{\mathcal P} $. Note that $(x^i(t),\tilde f^a(t))$ describe the local stochastic evolution on the base manifold $\tilde{\mathcal M}$ of the principal fiber bundle.

Based on the replacement of the initial coordinates with the adapted ones, the transformation of local random processes occurs as follows:
\begin{equation}
\eta_1^A(t)=F^A(Q^{\ast}{}^B(x^i(t)),a^{\alpha}(t)),\;\;\eta_2^a(t)=\bar D^a_c(a(t))\tilde f^c(t).
\label{phasespace}
\end{equation}
This transformation is the phase space transformation of the stochastic processes on which it is known that it  does not change the probabilities.

This means the following. If we restrict  the local semigroup 
\[
{\tilde U}_{\eta}(s,t) \phi _0 (p,v)=
{\rm E}_{s,(p,v)}\phi _0 (\eta_1 (t),\eta_2(t)),\,\,\,
s\leq t,\,\,\,\eta_1 (s)=p,\;\;\;\eta_2 (s)=v
\]
for the process $\eta (t)$  to the chart 
$({\mathcal{U}}_{(p,v)},\varphi ^{\tilde{\cal P}})$ with 
\[
\varphi ^{\tilde{\cal P}}(\eta (t))=\{{\eta}_1^{\varphi 
^{\tilde{\cal P}}}(t),{\eta}_2^{\varphi 
^{\tilde{\cal P}}}(t)\}\equiv\{\eta_1^A (t),\eta_2^a(t)\},
\]
 where this semigroup can be be represented as 
\[
 {\tilde U}_{\eta}(s,t) \phi _0 (p,v)={\rm E}_
 {s,\varphi ^{\tilde{\cal P}}(p,v)}
 \phi _0\left((\varphi ^{\tilde{\cal P}})^{-1}
 (\eta_1^{\varphi ^{\tilde{\cal P}}}(t),\eta_2^{\varphi ^{\tilde{\cal P}}}(t))\right),
\] 
\[
\,\,\,\eta_1^{\varphi ^{\tilde{\cal P}}}(s)=
 \varphi ^{\tilde{\cal P}}(p),\;\eta_2^{\varphi ^{\tilde{\cal P}}}(s)=
 \varphi ^{\tilde{\cal P}}(v),
\]
then as a result of the phase space transformation (\ref{phasespace}) of the local stochastic processes 
we get
\begin{equation}
{\tilde U}_{\eta}(s,t) \phi _0(p)={\rm E}_
{s,{\tilde{\varphi}}^{\tilde{\cal P}}(p,v)}
\phi _0\left(({\tilde{\varphi}}^{\tilde{\cal P}})^{-1}
({\zeta}^{{\tilde{\varphi}}^{\tilde{\cal P}}}(t))\right)=
{\rm E}_
{s,{\tilde{\varphi}}^{\tilde{\cal P}}(p,v)}
{\tilde{\phi _0}}
\left(
{\zeta}^{{\tilde{\varphi}}^{\tilde{\cal P}}}(t)\right),
\label{semigr_zeta}
\end{equation}
with $\left(
{\tilde{\varphi}}^{\tilde{\cal P}}
\right)^{-1}
=\left(\varphi ^{\tilde{\cal P}}\right)^{-1}
\circ (F,\bar D)$ and ${\tilde{\phi _0}}={\phi _0}\circ
({\tilde{\varphi}}^{\tilde{\cal P}})^{-1}$.

Therefore, changing the coordinates leads to the replacement of the measure used to calculate the expectation values. 
Now measure is  defined by  the probability
distribution of the local processes
${\zeta}^{{\tilde{\varphi}}^{\cal
P}}(t)=
(x^i(t),\tilde{f^a}(t), a^{\alpha}(t))$.
According to method developed in \cite{Dalecky_1,Dalecky_2}, to define the global process and the global semigroup, it is necessary that the local processes are consistent with each other on  overlapping of the charts. In our case, this consistency follows  from the transformation properties of local stochastic  equations. These equations are the Stratonovich equations.

\section{Stochastic differential equations for the transformed process}
 Local stochastic differential equations for the process $\zeta (t)$, which defines the integration measure in the evolution semigroup obtained as a result of the transformation of the process $\eta(t)$, can be derived by the It\^{o} differentiation formula together with using the initial stochastic differential equations (\ref{eta_1}) and (\ref{eta_2}). Since the adapted coordinates depend on the intial coordinates as $x^i(Q)$, $a^{\alpha}(Q)$ and $\tilde f^{a}(Q,f)$, we should differentiate (by It\^{o}) the following functions: $x^i(\eta_1(t))$, $a^{\alpha}(\eta_1(t))$ and $\tilde f^{a}(\eta_1(t),\eta_2(t))$. Let us consider, for example,  how 
the stochastic differential equation for the component $x^i(t)$ of the local process $\zeta(t)$ is obtained.

The It\^{o}  differential of $x^i(t)$ can be written as 
$$dx^i(t)=\frac{\partial x^i}{\partial Q^A}(\eta_1(t))d\eta_1^A(t)+\frac12\frac{\partial ^2 x^i}{\partial Q^A\partial Q^B}(\eta_1(t))d\eta^A_1(t)d\eta_1^B(t),$$
where 
$$\frac{\partial x^i}{\partial Q^A}(F(Q^{\ast}(x),a))={\check F}^B_A G^{\rm H}_{BC}(Q^{\ast}(x))Q^{\ast C}_m(x)h^{mi}(x)$$
in which ${\check F}^B_A\equiv F^B_A(F(Q^{\ast}(x),a),a^{-1})$ is the inverse matrix to the matrix $F^A_B$.

Using (\ref{eta_1}) on the right hand side of the expression for $dx^i(t)$ and performing the necessary transformations, we obtain
\begin{equation}
 dx^i(t)=\frac12 (\mu^2\kappa)b^i(x(t))dt +\mu\sqrt{\kappa}X^i_{\bar M}(Q^{\ast}(x(t)))dw^{\bar M}(t), 
\label{sde_x}
\end{equation}
where 
\[
 b^i=d^{-1/2}H^{-1/2}\frac{\partial}{\partial x^j}(d^{1/2}H^{1/2}h^{ij})+\underset{\scriptscriptstyle{(\gamma)}}{{\mathscr A}^{\mu}_n}  h^{ni}d^{-1/2}
H^{-1/2}\frac{\partial}{\partial \tilde f^b}(d^{1/2}H^{1/2}K^b_{\mu})
\]
with $K^b_{\alpha '}=(\bar J_{\alpha '})^b_c\,\tilde f^c$, and $X^i_{\bar M}$ is determined by the following equality:
$X^i_{\bar M}(Q^{\ast}(x(t)))=\frac{\partial x^i}{\partial Q^A}{\rm {\mathcal X}}^A_{\bar M}(\eta_1(t))$, $\eta_1^A=F^A(Q^{\ast}(x),a)$, which leads to 
\[
 X^i_{\bar M}(Q^{\ast}(x))=T^i_D(Q^{\ast}(x))N^D_C(Q^{\ast}(x))\mathcal X^c_{\bar M}(Q^{\ast}(x)).
\]
This expression is written in terms of two projection operators. One of them, $N^D_C=\delta ^D_C-K^C_{\mu}\Lambda ^{\mu}_C$, is the projection operator onto the subspace which is orthogonal to the Killing vector field space. In our case, this operator is taken on the submanifold $\Sigma$, that is, $N^D_C=N^D_C(Q^{\ast}(x)).$ 

The second operator,
$$T^i_D=(P_{\bot})^B_D(Q^{\ast}(x))G^{\rm H}_{BL}(Q^{\ast}(x))Q^{\ast L}_m(x) h^{mi}(x),$$ with the properties $T^i_DQ^{\ast D}_j=\delta^i_j$ and $T^i_DQ^{\ast B}_i=(P_{\bot})^B_D$, is defined by using the projection operator $P_{\bot}$ on the tangent plane to the submanifold $\Sigma$ of the manifold $\mathcal P$:
\[
(P_{\perp})^{A}_{B}=\delta ^{A}_{B}-{\chi}^{\alpha}_{B}
(\chi \chi ^{\top})^{-1}{}^{\beta}_{\alpha}(\chi ^
{\top})^{A}_{\beta}.
\]
Here $(\chi ^{\top})^{A}_{\beta}$ is a transposed matrix to the
matrix
$\chi ^{\nu}_{B}$:
\[
(\chi ^{\top})^{A}_{\mu}=G^{AB}{\gamma}_
{\mu \nu}\chi ^{\nu}_{B},\,\,\, {\gamma}_{\mu \nu}
=K^{A}_{\mu}G_{AB}K^{B}_{\nu}.
\]
The  projection operators have the following properties:
\[
(P_{\perp})^{\tilde A}_{B}N^{C}_{\tilde A}=
(P_{\perp})^{C}_{B},\,\,\,\,\,\,\,\,\,N^{\tilde A}_
{B}(P_{\perp})^{C}_{\tilde A}=N^{C}_{B}.
 \]

Note that from the above definition of $X^A_{\bar M}$ it follows that $\sum_{\bar M}
X^i_{\bar  M}X^j_{\bar M}=h^{ij}(x),$ where $h^{ij}$ is the metric in the base space $M$ of  the principal fiber bundle ${\rm P}(M,\mathcal G)$.

Taking into account that $ \tilde f^a=\tilde f^a(Q,f), a^{\alpha}=a^{\alpha}(Q),$ the local stochastic differential equations for the stochastic variables $\tilde f^a(t)$ and $a^{\alpha}(t)$   can be derived in the same way as it was done for $x^i(t)$. They are as follows:
\begin{equation}
 d\tilde f^a(t)=\frac12(\mu^2\kappa)b^a(x(t),\tilde f(t))dt +\mu\sqrt{\kappa} X^a_{\bar M}(Q^{\ast}(x(t)),\tilde f(t))dw^{\bar M}(t)+X^a_{\bar b}
dw^{\bar b}(t)),
\label{sde_f} 
\end{equation}
where the function $X^a_{\bar M}=N^a_A\mathcal X^A_{\bar M}(Q^{\ast}(x))$,  $X^a_{\bar b}=\mathcal X^a_{\bar b}$ and
\begin{eqnarray*}
b^a&=&d^{-1/2}H^{-1/2}\frac{\partial}{\partial x^j}(d^{1/2}H^{1/2}\underset{\scriptscriptstyle{(\gamma)}}{{\mathscr A}^{\mu}_m}  h^{mj})K^a_{\mu}\\
&+&(G^{ab}+G^{AB}N^a_AN^b_B)d^{-1/2}H^{-1/2}\frac{\partial}{\partial \tilde f^b}(d^{1/2}H^{1/2})+\frac{\partial}{\partial \tilde f^b}(G^{AB}N^a_AN^b_B).
\end{eqnarray*}
(The last term of the previous expression is equal to $2G^{AB}{\Lambda}^{\mu}_A{\Lambda}^{\nu}_B(\bar J_{\mu})^a_b(\bar J_{\nu})^b_c\tilde f^c$.)

The stochastic differential equation for $a^{\alpha}(t)$ is given by
\begin{equation}
da^{\alpha}(t)=\frac12(\mu^2\kappa)b^{\alpha}dt+\mu\sqrt{\kappa}X^{\alpha}_{\bar M}(Q^{\ast}(x(t)),a(t))dw^{\bar M}(t),
\label{sde_a}
\end{equation}
where $X^{\alpha}_{\bar M}={\Lambda}^{\mu}_A(Q^{\ast}(x))\bar v^{\alpha}_{\mu}(a))\mathcal X^A_{\bar M}(Q^{\ast}(x))$ and 
\begin{eqnarray*}
 b^{\alpha}&=&-d^{-1/2}H^{-1/2}\frac{\partial}{\partial x^j}(d^{1/2}H^{1/2}h^{nj}\,\underset{\scriptscriptstyle{(\gamma)}}{{\mathscr A}^{\beta}_n})\bar v ^{\alpha}_{\beta}+G^{BC}{\Lambda}^{\alpha'}_B{\Lambda}^{\beta'}_C \bar L_{\beta'}\bar v ^{\alpha}_{\alpha'}\\
&-&G^{EC}{\Lambda}^{\beta}_E{\Lambda}^{\mu}_C\,d^{-1/2}H^{-1/2}\frac{\partial}{\partial \tilde f^b}(d^{1/2}H^{1/2}K^b_{\mu})\bar v ^{\alpha}_{\beta}.
\end{eqnarray*}

The coefficients $b^i$, $b^a$ and $ b^{\alpha}$ of the stochastic differential equations (\ref{sde_x}), (\ref{sde_f}), (\ref{sde_a}) are obtained  as a result of the transformations of the original equations (\ref{eta_1}) and (\ref{eta_2}). Note that they are equal to the coefficients  at first derivatives in the differential generator of the local evolution semigroup associated with the local process $\zeta(t)$. This differential generator can be derived by the standard methods from the stochastic theory and coinsides with the operator obtained from the initial differential operator $\frac12({\triangle}_{\mathcal P}+{\triangle}_{\mathcal V})$ after  the transition to the adapted coordinates.

Also we note that in obtained stochastic differential equations the diffusion terms are actually determined  up to an arbitrary orthogonal transformations \cite{McKean}. But this does not affect the measure in the path integral, since it can be compensated by the orthogonal transformations of the Wiener processes:  $(w^{\tilde A})^{'}=
O^{\tilde A}_{\tilde B}w^{\tilde B}.$ This transformations keep the measure invariant.

Next we will use stochastic differential equations to derive a nonlinear filtering stochastic differential equation of our problem. 
This equation will allow us to factorize the measure in the path integral, which represents a local evolution semigroup.

But before that, the local stochastic differential equations we obtained must be slightly transformed. The reason for this is that they must have a certain dependence on the diffusion coefficients. Namely, the form of the equations should be as follows:
\begin{eqnarray}
 dx^i &=& \frac12(\mu^2\kappa) b^i dt+\mu\sqrt{\kappa}\tilde X^i_{\bar m}\,d\tilde w^{\bar m}
\nonumber\\
 d\tilde f ^a &=& \frac12(\mu^2\kappa) b^a dt+ \mu\sqrt{\kappa}(\tilde X^a_{\bar m}\,d\tilde w^{\bar m}+\tilde X^a_{\bar b}\,d\tilde w^{\bar b})
\nonumber\\
 da^{\alpha} &=& \frac12(\mu^2\kappa) b^{\alpha} dt+\mu\sqrt{\kappa}(\tilde X^{\alpha}_{\bar m}\,d\tilde w^{\bar m}+\tilde X^{\alpha}_{\bar \beta}\,d\tilde w^{\bar \beta}+\tilde X^{\alpha}_{\bar b}\,d\tilde w^{\bar b})
\label{sde_filtr}
\end{eqnarray}
Now the Wiener processes $\tilde w^{\bar m},\;\tilde w^{\bar b},\;\tilde w^{\bar \beta}$ are independent of each other.
We see that to get (\ref{sde_filtr}) from the previous stochastic differential equations (\ref{sde_x}), (\ref{sde_f}), (\ref{sde_a}), it will not be enough if we only replace  $X^i_{\bar M}dw^{\bar M}$ with $X^i_{\bar m}dw^{\bar m}+X^i_{\bar \alpha}dw^{\bar \alpha}$, and similarly other diffusion terms.

The coefficients of the previous stochastic differential equations are replaced in accordance with \cite{Lipcer,Pugachev} as follows:
\begin{eqnarray*}
&&X^i_{\bar m}dw^{\bar m}+X^i_{\bar \alpha}dw^{\bar \alpha}=
{\tilde X}^i_{\bar m}d{\tilde w}^{\bar m},
\nonumber\\
&&X^a_{\bar m}dw^{\bar m}+X^a_{\bar \beta}dw^{\bar \beta}+X^a_{\bar b}dw^{\bar b}=
\tilde X^a_{\bar m}\,d\tilde w^{\bar m}+\tilde X^a_{\bar b}\,d\tilde w^{\bar b}
\nonumber\\
&&X^{\alpha}_{\bar m}dw^{\bar m}+X^{\alpha}_{\bar \beta}dw^{\bar
\beta}
=\tilde X^{\alpha}_{\bar m}\,d\tilde w^{\bar m}+\tilde X^{\alpha}_{\bar \beta}\,d\tilde w^{\bar \beta}+\tilde X^{\alpha}_{\bar b}\,d\tilde w^{\bar b}.
\nonumber\\
\end{eqnarray*}
This replacement can be achieved with the help of the orthogonal transformation of the Wiener
processes $w^{\bar M}$ and $w^{\bar b}$. So, such a replacement has not change the path integral measure.

The main requirement imposed on diffusion coefficients is that the local stochastic process with the stochastic differential  equations (\ref{sde_filtr}) must has the same differential generator of the local evolution semigroup as the process gaverned by the previous stochastic equations. 
This requirement leads to the equations for determining the diffusion coefficients:
\begin{eqnarray*}
 &&1.\;\;{\tilde X}^i_{\bar m}{\tilde X}^j_{\bar m}=h^{ij},
\nonumber\\
&&2.\;\;{\tilde X}^i_{\bar m}{\tilde X}^{a}_{\bar m}=\underset{\scriptscriptstyle{(\gamma)}}{{\mathscr A}^{\mu}_m} K^a_{\mu}h^{mi},
\nonumber\\
&&3.\;\;\tilde X^a_{\bar m}\tilde X^b_{\bar m}+\tilde X^a_{\bar b}\tilde X^b_{\bar b}=G^{AB}N^a_AN^b_B+G^{ab},
\nonumber\\
&&4.\;\;{\tilde X}^i_{\bar m}{\tilde X}^{\alpha}_{\bar m}=-h^{ni}\underset{\scriptscriptstyle{(\gamma)}}{{\mathscr A}^{\beta}_n}\bar v^{\alpha}_{\beta},
\nonumber\\
&&5.\;\;\tilde X^a_{\bar m}{\tilde X}^{\alpha}_{\bar m}+\tilde X^a_{\bar b}{\tilde X}^{\alpha}_{\bar b}=-\Bigl(\gamma^{\mu\nu}+h^{ij}\underset{\scriptscriptstyle{(\gamma)}}{{\mathscr A}^{\mu}_i}\underset{\scriptscriptstyle{(\gamma)}}{{\mathscr A}^{\nu}_j}\Bigr)K^a_{\mu}\bar v^{\alpha}_{\nu},
\nonumber\\
&&6.\;\;\tilde X^{\alpha}_{\bar m}\tilde X^{\beta}_{\bar m}+\tilde X^{\alpha}_{\bar a}\tilde X^{\beta}_{\bar a}+\tilde X^{\alpha}_{\bar \varepsilon}\tilde X^{\beta}_{\bar \varepsilon}=\Bigl(\gamma^{\mu\nu}+h^{ij}\underset{\scriptscriptstyle{(\gamma)}}{{\mathscr A}^{\mu}_i}\underset{\scriptscriptstyle{(\gamma)}}{{\mathscr A}^{\nu}_j}\Bigr)\bar v^{\alpha}_{\mu}\bar v^{\beta}_{\nu}.
\end{eqnarray*}
The first equation means that ${\tilde X}^i_{\bar m}=(h^{ij})^{1/2}$. Then from the second equation we can find that ${\tilde X}^{a}_{\bar m}={\tilde X}^k_{\bar m}\underset{\scriptscriptstyle{(\gamma)}}{{\mathscr A}^{\mu}_k} K^a_{\mu}$. Substituiting such representations for diffusion coefficients in the third equation, we obtain that  
$$\tilde X^a_{\bar b}\tilde X^b_{\bar b}=(\gamma^{\alpha \beta}K^a_{\alpha}K^b_{\beta}+G^{ab}).$$
This was obtained  using the equality $G^{AB}N^a_AN^b_B=(\gamma^{\alpha \beta}+h^{ij}\underset{\scriptscriptstyle{(\gamma)}}{{\mathscr A}^{\alpha}_i}\underset{\scriptscriptstyle{(\gamma)}}{{\mathscr A}^{\beta}_j})K^a_{\alpha}K^b_{\beta}.$
So we can write that  $\tilde X^a_{\bar b}=(\gamma^{\alpha \beta}K^a_{\alpha}K^b_{\beta}+G^{ab})^{1/2}$.

Next, we  find the representation for ${\tilde X}^{\alpha}_{\bar b}$. To do this, we first replace
${\tilde X}^{a}_{\bar m}$ and         ${\tilde X}^{\alpha}_{\bar m}$ of the fifth equation for their already known representations.  After simplifying the resulting expression, we get
\[
 {\tilde X}^{a}_{\bar b}{\tilde X}^{\alpha}_{\bar b}=-\gamma^{\mu \beta}K^a_{\mu}\bar v^{\alpha}_{\beta}.
\]
Assuming that ${\tilde X}^{\alpha}_{\bar b}={\tilde X}^{c}_{\bar b}Z^{\alpha}_c$, we obtain the equation for $Z^{\alpha}_c$:
\[
 (\gamma^{\alpha \beta}K^a_{\alpha}K^c_{\beta}+G^{ac})Z^{\alpha}_c=-\gamma^{\mu \beta}K^a_{\mu}\bar v^{\alpha}_{\beta}, \;\;\rm{or}
\]
\[
 R^{ac}Z^{\alpha}_c=-\gamma^{\mu \beta}K^a_{\mu}\bar v^{\alpha}_{\beta}.
\]
Therefore, we can write that $Z^{\alpha}_c=(R^{-1})_{ca}(-K^a_{\mu}\gamma^{\mu \beta}v^{\alpha}_{\beta})$. Therefore, this solution implies that ${\tilde X}^{a}_{\bar b}={\tilde X}^{c}_{\bar b}Z^{\alpha}_c$.

In a similar way, one can find a representation for the diffusion coefficient 
 $\tilde X^{\alpha}_{\bar \beta}$ using the  sixth equation:  
\[
 \tilde X^{\alpha}_{\bar \beta}=\bar v^{\alpha}_{\alpha'}\tilde Y^{\alpha'}_{\bar \beta}\;\;{\rm with}\;\;
\tilde Y^{\alpha'}_{\bar \beta}=(\gamma^{\alpha'\beta'}-\gamma^{\alpha'\mu}K^a_{\mu}(R^{-1})_{ab}K^b_{\nu}\gamma^{\nu\beta'})^{1/2}.
\]

The stochastic process, which is solution of the stochastic differential equations (\ref{sde_filtr}),  will  be denoted by $\tilde\zeta(t)$.
The differential generator for the semigroup related with this process is the same as for the process obtained from the solution equations (\ref{sde_x}), (\ref{sde_f}), (\ref{sde_a}). 
The process $\tilde\zeta(t)$ generate the same path integral measure as the process $\zeta(t)$.
This means that we could transform the process $\eta(t)$ to our process $\tilde\zeta(t)$ from the very beginning, and therefore the equality (\ref{semigr_zeta}) also extends to the semigroup associated with $\tilde\zeta(t)$.

The global semigroup for the process $\tilde\zeta(t)$ is obtained as a limit of the local semigroups based on the local process 
${\tilde\zeta}^{{\tilde{\varphi}}^{\tilde{\cal P}}}(t)$:
\begin{equation}
\psi _{t_b}(p_a,v_a,t_a)=
{\lim}_q {\tilde U}_{{\tilde\zeta}^{{\tilde{\varphi}}^{\tilde{\cal P}}}}(t_a,t_1)\cdot\dots\cdot
{\tilde U}_{{\tilde\zeta}^{{\tilde{\varphi}}^{\tilde{\cal P}}}}
(t_{n-1},t_b) 
{\tilde \phi} _0(x_a,\tilde f _a, \theta _a),
\label{glob_semigr_zeta}
\end{equation}
where by ${\tilde U}_{{\tilde\zeta}^{{\tilde{\varphi}}^{\tilde{\cal P}}}}$ we denote 
\begin{eqnarray}
&&{\tilde U}_{{\tilde\zeta}^{{\tilde{\varphi}}^{\tilde{\cal P}}}}(s,t) 
{\tilde \phi} (x_0,\tilde f_0,\theta _0)={\rm E}_
{s,(x_0,\tilde f_0,\theta _0)}
\tilde{\phi}(x(t),\tilde f(t),a(t)),\;\;
\nonumber\\
&&\,\,{\rm with}\;\;x(s)=x_0,\,\tilde f(s)=\tilde f_0,\,a(s)=\theta _0.
\label{local_semigr_zeta}
\end{eqnarray}

\section{Factorization  of the path integral measure}

Factorization of the  path integral measure in the path integrals for dynamical system with symmetry was considered in our works \cite{Storchak_1,Storchak_11,Storchak_1,Storchak_2}. There, as well as in \cite{Elworthy}, it was shown that this procedure is based on the use of stochastic
differential equation from the theory of nonlinear filtering.
\cite{Lipcer, Pugachev}. 

 This equation describes the evolution of the conditional mathematical expectation of the signal process 
 with respect to the $\sigma $--algebra generated by an observable process. In our case, we consider the stochastic process $a(t)$ as a signal process. As an observable process, there will be a process consisting of $x(t)$ and $\tilde f(t)$. Note that that this observable process describes the stochastic evolution on the base space $\tilde{\mathcal M}$ of the principal fiber bundle 
$\rm {P}(\tilde{\mathcal M},\mathcal G)$.

The properties of the conditional mathematical expectation of the Markov processes allow us to express each local semigroup (\ref{local_semigr_zeta}) of the global semigroup (\ref{glob_semigr_zeta}) as follows:
  \begin{equation}
{\tilde U}_{{\tilde\zeta}^{{\tilde{\varphi}}^{\tilde{\cal P}}}}(s,t) {\tilde
\phi} (x_0,\tilde f_0,\theta _0)=
{\rm E}
\Bigl[{\rm E}\Bigl[\tilde{\phi }(x(t),\tilde f(t),a(t))\mid
(
{\cal F}_{(x,\tilde f)})_{s}^{t}\Bigr]\Bigr].
\label{loc_cond_expect}
\end{equation}

The nonlinear filtering equation is derived for 
the conditional mathematical expectation  on  the right--hand
 side of (\ref{loc_cond_expect}),
 \[
 \hat{\widetilde{\phi }}((x(t),\tilde f(t))\equiv 
 {\rm E}\Bigl[\tilde{\phi }(x(t),\tilde f(t),a(t))\mid (
 {\cal F}_{(x,\tilde f)})_{s}^{t}\Bigr].
\]

In  \cite{Lipcer,Pugachev}, the nonlinear filtering stochastic
 differential equation for 
\[
 \hat f(t)=\rm E[f(Z,t)|Y^t_{t_0}]
\]
is defined as follows:
\begin{eqnarray} 
d\hat f(t)&=&\rm E[f_t+f_z\varphi +\frac12 f_{zz}(X X^T)|Y^t_{t_0}]dt
\nonumber\\
&&\!\!\!+\rm E\Bigl[f(\varphi_1-\hat \varphi_1)+f_z(X X^T_1)|Y^t_{t_0}\Bigr](X_1 X^T_1)^{-1}(dY-\hat \varphi_1 dt)
\label{eq_filtr_pugach}
\end{eqnarray}

We can regard the processes $x^i(t)$ and $\tilde f^a(t)$ as a process  $Y_t$ of the above equation:
\begin{equation}
 {dx^i\choose d\tilde f^a}=\frac12(\mu^2\kappa) {b^i\choose b^a}dt+\mu \sqrt{\kappa}{\tilde X^i_{\bar m}\;\; 0\choose {\tilde X^a_{\bar m}\;\; \tilde X^a_{\bar b}}} {d\tilde w^{\bar m}\choose d\tilde w^{\bar b}}.
\label{sde_x_f}
\end{equation}
The role of the process $Z_t$ is played by the stochastic process $a^{\alpha}(t)$:
\[
 da^{\alpha}=\frac12(\mu^2\kappa)b^{\alpha} dt +\mu \sqrt{\kappa}(\tilde X^{\alpha}_{\bar m}\;\;\tilde X^{\alpha}_{\bar b}\;\;\tilde X^{\alpha}_{\bar \beta})
\begin{pmatrix}
 d\tilde w^{\bar m} \\ d\tilde w^{\bar b} \\ d\tilde w^{\bar \beta}
\end{pmatrix}.
\]
With  such identifications one can find the explicit representations for the  terms of the equation (\ref{eq_filtr_pugach}). The resulting nonlinear stochastic differential equation in our case is as follows:
\begin{eqnarray}
 &&d \hat{\tilde\phi}(x(t),\tilde f(t))=\mu^2\kappa\Bigr\{-\frac12\Bigl[d^{-1/2}H^{-1/2}\frac{\partial}{\partial x^j}(d^{1/2}H^{1/2}h^{nj}\,\underset{\scriptscriptstyle{(\gamma)}}{{\mathscr A}^{\beta}_n})
\nonumber\\
&&+G^{EC}{\Lambda}^{\beta}_E{\Lambda}^{\mu}_C\,d^{-1/2}H^{-1/2}\frac{\partial}{\partial \tilde f^b}(d^{1/2}H^{1/2}K^b_{\mu})
\Bigr]\rm E\bigl[\bar L_{\beta}{\tilde\phi}(x,\tilde f,a)|(\mathcal F_{x,\tilde f})^t_s\bigr]
\nonumber\\
&&+\frac12(G^{BC}{\Lambda}^{\alpha'}_B{\Lambda}^{\beta'}_C)\rm E\bigl[\bar L_{\alpha'}\bar L_{\beta'}{\tilde\phi}(x,\tilde f,a)|(\mathcal F_{x,\tilde f})^t_s\bigr]\Bigr\}dt
\nonumber\\
&&-\mu\sqrt{\kappa}\,\underset{\scriptscriptstyle{(\gamma)}}{{\mathscr A}^{\beta}_n}\tilde X^n_{\bar m}\rm E\bigl[\bar L_{\beta}{\tilde\phi}(x,\tilde f,a)|(\mathcal F_{x,\tilde f})^t_s\bigr]d\tilde w^{\bar m}
\nonumber\\
&&-\mu\sqrt{\kappa}\,\tilde {\mathscr A}^{\beta}_a\tilde X^a_{\bar b}\rm E\bigl[\bar L_{\beta}{\tilde\phi}(x,\tilde f,a)|(\mathcal F_{x,\tilde f})^t_s\bigr]d\tilde w^{\bar b}
\label{eq_filtr}
\end{eqnarray}

This equation can be simplified if we apply the Peter-Weyl theorem to the function $\tilde\phi$.
 This can be done since $\tilde\phi$ depends on a group variable $a$, and thus, it is a function  on a group $\mathcal G$.
By this theorem
$$\tilde \phi(x,\tilde f,a)=\sum_{\lambda,p,q}c^{\lambda}_{pq}(x,\tilde f)D^{\lambda}_{pq}(a),$$
 where $D^{\lambda}_{pq}(a)$\footnote{Now we have introduce another notation for the matrix elements  of an irreducible representation in order to distiguish them from those   that were used earlier.} are 
    the matrix elements  of an irreducible representation
    $T^{\lambda}$ of a group $\cal G$:
 $\sum_qD_{pq}^\lambda(a)D_{qn}^\lambda (b)=D_{pn}^\lambda (ab)$.
 
From the properties of the conditional mathematical expectations, it follows that
\begin{eqnarray*}
\rm E\bigl[\tilde \phi(x(t),\tilde f(t),a(t))|(\mathcal F_{x,\tilde f})^t_s\bigr]&=&\sum_{\lambda,p,q}c^{\lambda}_{pq}(x(t),\tilde f(t))\,{\rm E}\bigl[D^{\lambda}_{pq}(a(t))|(\mathcal F_{x,\tilde f})^t_s\bigr]\\
&\equiv& \sum_{\lambda,p,q}c^{\lambda}_{pq}(x(t),\tilde f(t))\,\hat D^{\lambda}_{pq}(x(t),\tilde f(t))
\end{eqnarray*}
where
\[
c_{pq}^\lambda (x(t),\tilde f(t))=d^\lambda \int_{\mathcal
G}\tilde{\varphi }
(Q^{*}(t),\theta ) 
{\bar D}_{pq}^\lambda (\theta )d\mu (\theta ),
\]
$d^{\lambda}$ is a dimension of an irreducible representation
and 
$d\mu (\theta )$ is a normalized\\ ($\int_{\mathcal G}d\mu
(\theta )=1$)
invariant Haar measure on a group $\mathcal G$.

It can be shown that the  conditional mathematical expectation $\hat{D}_{pq}^\lambda$
satisfies  the linear matrix equation:
\begin{eqnarray}
&&d\hat{D}_{pq}^\lambda =\Gamma _1^\mu \,\,(J_\mu )_{pq^{\prime
}}^\lambda \hat{D}_{q^{\prime }q}^\lambda dt
+\Gamma _2^{\mu \nu }\,\,(J_\mu )_{pq^{\prime }}^\lambda (J_\nu
)_{q^{\prime }q^{\prime \prime }}^\lambda \hat{D}_{q^{\prime \prime
}q}^\lambda dt
\nonumber\\
&&-(J_\mu )_{pq^{\prime }}^\lambda \hat{D}_{q^{\prime }q}^\lambda\,\bigl(
\underset{\scriptscriptstyle{(\gamma)}}{{\mathscr A}^{\beta}_n}\tilde X^n_{\bar m}d\tilde w^{\bar m}+\tilde {\mathscr A}^{\beta}_a\tilde X^a_{\bar b}d\tilde w^{\bar b}\bigr).
\label{eq_filtr_D}
\end{eqnarray}
In this equation,   $(J_\mu )_{pn}^\lambda$ are the infinitesimal generators
of the representation $D^{\lambda}(a)$. They are determined as  
$(J_\mu )_{pq}^\lambda \equiv (\frac{\partial D_{pq}^\lambda
(a)}{\partial
a^\mu })|_{a=e}$, so that
\[
\bar{L}_\mu D_{pq}^\lambda (a)=\sum_{q^{\prime }}(J_\mu )_{pq^{\prime
}}^\lambda D_{q^{\prime }q}^\lambda (a).
\] 
 Regarding the   explicit form  of the coefficients  $\Gamma
_1^{\mu}$ and $\Gamma _2^{\mu \nu}$ we notice that they  can be easily  obtained from (\ref{eq_filtr}).   
 Also note that in notation of  the conditional expectations
  $\hat{D}_{pq}^\lambda (x(t),\tilde f(t))$ we omit the existing dependencies
  on the initial points:
  $x_0=x(s)$, $\tilde f^a_0=\tilde f^a(s)$ and $\theta^{\alpha}_0=a^{\alpha}(s)$.

Based on the  approach to solving linear matrix stochastic differential equations, which was developed 
  in \cite{Dalmulti,Stroock}, we write the solution of our equation (\ref{eq_filtr_D}) as follows:
 \begin{equation}
\hat{D}_{pq}^\lambda (x(t),\tilde f(t))=(\overleftarrow{\exp })_{pn}^\lambda
(x(t),\tilde f(t),t,s)\,{\rm E}\bigl[D_{nq}^\lambda (a(s))\mid ({\cal F}
_{x,\tilde f})_{s}^t\bigr],
\label{solut_eq_filtr_D}
\end{equation}
where 
\begin{eqnarray}
&&(\overleftarrow{\exp })_{pn}^\lambda (x(t),\tilde f(t),t,s)=
\overleftarrow{\exp }%
\int_{s}^t\Bigl\{{\mu}^2\kappa\Bigl[\frac 12d^{\alpha \nu
}(x(u),\tilde f(u))(J_\alpha
)_{pr}^\lambda (J_\nu )_{rn}^\lambda \Bigr.\Bigr.
\nonumber\\
&&-\Bigl.\Bigl.\frac 12\frac 1{\sqrt{d\, H}}\frac \partial {\partial x^k}\left( \sqrt{d\, H}%
h^{km}\underset{\scriptscriptstyle{(\gamma)}}{{\mathscr A}_m^\nu}(x(u)) \right) (J_\nu )_{pn}^\lambda  
\nonumber\\
&&-\frac12(G^{EC}\Lambda^{\nu}_E\Lambda^{\mu}_C)\frac 1{\sqrt{d\, H}}\frac \partial {\partial \tilde f^b}\left(\sqrt{d\, H}K^b_{\mu}\right)(J_\nu )_{pn}^\lambda \Bigr]du
\nonumber\\
&&-\mu\sqrt{\kappa}\Bigl[\underset{\scriptscriptstyle{(\gamma)}}{{\mathscr A}^{\nu}_k}(x(u))\tilde X^k_{\bar m}(u)(J_\nu )_{pn}^\lambda d\tilde w^{\bar m}(u)+\tilde {\mathscr A}^{\nu}_a\tilde X^a_{\bar b}(u)(J_\nu )_{pn}^\lambda d\tilde w^{\bar b}(u)\Bigr]\Bigr\}
\nonumber\\
&&(H,\tilde {\mathscr A}^{\nu}_a\;{\rm depend}\,{\rm on}\;x(u)\;{\rm and}\,\tilde f(u)) 
\label{multipl_exp}
\end{eqnarray}
is a multiplicative stochastic integral. This integral is defined as a
limit of
the sequence of time--ordered multipliers that have been
obtained as a
result of breaking of a time interval $[s,t]$, $[s=t_0\le t_1
\ldots \le t_n=t]$.
In (\ref{multipl_exp}), the time order of these
multipliers is indicated by the arrow directed 
to the multipliers given at greater times.
 
Thus, (\ref{solut_eq_filtr_D}) and (\ref{multipl_exp}) give us   the necessary representation for $\hat{D}_{pq}^\lambda$. Therefore,  the local semigroup  (\ref{loc_cond_expect}) can now be rewritten  in the following form:
\begin{equation}
{\tilde U}_{{\tilde\zeta}^{{\tilde{\varphi}}^{\tilde{\cal P}}}}(s,t) {\tilde \phi} (x_0,\tilde f_0,\theta _0)
=\sum_{\lambda ,p,q,q^{\prime }}{\rm E}
\bigl[
 c_{pq}^\lambda (x(t),\tilde f(t))
(\overleftarrow{\exp })_{pq^{\prime }}^\lambda 
(x(t),t,s)\bigr] D_{q^{\prime}q}^\lambda (\theta _0),
\label{local_semigr_exp}
\end{equation}
where it was  
 taken into account that 
\[
{\rm E}\bigl[D_{nq}^\lambda (a(s))\mid ({\cal F}_{x,\tilde f})_{s}^t\bigr]
=D_{nq}^\lambda (a(s))=D_{nq}^\lambda (\theta _0).
\]

Then the global semigroup (\ref{glob_semigr_zeta}), which is formed as  a superposition of the local semigroups similar to (\ref{local_semigr_exp}), is written symbolically as follows:
\begin{eqnarray}
{\psi}_{t_b}(p_a,v_a,t_a)
&=&\sum_{\lambda ,p,q,q^{\prime }}{\rm E}
\bigl[
 c_{pq}^\lambda (\xi(t_b))
(\overleftarrow{\exp })_{pq^{\prime }}^\lambda 
(\xi(t),t_b,t_a)\bigr] D_{q^{\prime}q}^\lambda (\theta _{a}),
\nonumber\\
&&\;\;\;\;\;\;\;\;\;(\xi(t_a)=\pi' \circ (p_a,v_a)),
\label{glob_semigr_ksi}
\end{eqnarray}
where the global process $\xi (t)=(\xi_1(t),\xi_2(t))$ is defined  on the manifold ${\tilde{\cal
M}}={\cal P}\times_{\cal G}{\cal \mathcal V}$.
The local stochastic evolution of  the process
$\xi (t)$ is given by the solution of the stochastic equations (\ref{sde_x_f}).

Thus, we have transformed our original path integral (\ref{orig_path_int}), and now it is represented by the right-hand side of (\ref{glob_semigr_ksi}), i.e., as the sum of the matrix semigroups (the path integrals)  on the manifold $\tilde{\mathcal M}$.

The differential generator (the Hamilton operator) of these matrix semigroups is 
\begin{eqnarray}
&&\frac 12\mu ^2\kappa \Bigl\{\Bigl[\triangle _{\tilde M}
+h^{ni}\Bigl(\frac 1{\sqrt{d}%
}\frac {\partial (\sqrt{d}\,)}
{\partial x^n}+\underset{\scriptscriptstyle{(\gamma)}}{{\mathscr A}^{\nu}_n}\frac 1{\sqrt{d}%
}\frac {\partial (\sqrt{d}K^b_{\nu}\,)}
{\partial \tilde f^b}\Bigr)\frac \partial
{\partial x^i}
\nonumber\\
&&+\Bigl(h^{mi}\underset{\scriptscriptstyle{(\gamma)}}{{\mathscr A}^{\mu}_m}K^a_{\mu}\frac 1{\sqrt{d}%
}\frac {\partial (\sqrt{d}\,)}{\partial  x^i}+(G^{ab}+G^{AB}N^a_AN^b_B)\frac 1{\sqrt{d}%
}\frac {\partial (\sqrt{d}\,)}
{\partial \tilde f^b}\Bigr)\frac \partial
{\partial \tilde f^a}
\Bigr](I^\lambda )_{pq}
\nonumber\\
&&-2h^{ni}\underset{\scriptscriptstyle{(\gamma)}}{{\mathscr A}^{\beta}_n} (J_\beta )_{pq}^\lambda
\frac \partial {\partial x^i}-2h^{nk}\underset{\scriptscriptstyle{(\gamma)}}{{\mathscr A}^{\beta}_n} \underset{\scriptscriptstyle{(\gamma)}}{{\mathscr A}^{\mu}_k}K^a_{\mu}(J_\beta )_{pq}^\lambda \frac \partial {\partial \tilde f^a}
\bigr.
\nonumber\\
&&-2({\gamma}^{\alpha\beta}K^a_{\alpha}K^b_{\beta}+G^{ab})\tilde {\mathscr A}^{\beta}_a(J_\beta )_{pq}^\lambda \frac \partial {\partial \tilde f^b}
\bigr.
\nonumber\\
&&-\bigl.\Bigl[\frac 1{\sqrt{dH}}\frac
\partial {\partial x^i}\left( \sqrt{dH}h^{ni}\underset{\scriptscriptstyle{(\gamma)}}{{\mathscr A}^{\beta}_n}\right)  
+(G^{EC}\Lambda^{\beta}_E\Lambda^{\mu}_C)\frac 1{\sqrt{dH}}
\frac {\partial (\sqrt{dH}K^b_{\mu}\,)}{\partial  \tilde f^b}\Bigr](J_\beta )_{pq}^\lambda 
\nonumber\\
&&+({\gamma }^{\alpha \beta }+h^{ij}\underset{\scriptscriptstyle{(\gamma)}}{{\mathscr A}^{\alpha}_i}\underset{\scriptscriptstyle{(\gamma)}}{{\mathscr A}^{\beta}_j})(J_\alpha
)_{pq^{\prime }}^\lambda (J_\beta )_{q^{\prime }q}^\lambda \Bigr\}.
\label{dif_gen_ksi}
\end{eqnarray} 
(Here  $(I^\lambda )_{pq}$ is a unity matrix.) 

This operator  acts in the space of the sections 
$\Gamma ({\tilde{\cal M}},V^{*})$ of the associated co-vector 
bundle (we consider the backward equation) 
${\cal E}^{*}={\tilde{\cal P}}\times _{\cal G}V^{\ast}_\lambda $, $ {\tilde{\cal P}}=\cal P\times\cal \mathcal V$.
The scalar product in the space of the sections of 
the associated co-vector  bundle is given by the following form:
\begin{equation}
(\psi _n,\psi _m)=\int_{\tilde{\cal M}}\langle \psi _n,\psi _m{\rangle}_
{V^{\ast}_{\lambda}}
 \sqrt{d(x,\tilde f)}dv_{\tilde{\cal M}}(x,\tilde f),
\label{33}
\end{equation}
 where $dv_{\tilde{\cal M}}(x,\tilde f)=\sqrt{H(x,\tilde f)}dx^1...dx^{n_{\cal M}}d\tilde f^1...\tilde f^{n_{\cal \mathcal V}}$ 
is the Riemannian volume element on the manifold ${\tilde{\cal M}}$. 

Equality (\ref{glob_semigr_ksi}) is in fact  the relation between the path integrals. On the left side, we have the path integral from (\ref{orig_path_int}) representing the solution of equation (1).  An expression with matrix semigroups (path integrals) on the reduced space is on the right. The resulting equality can be reversed.  In this case, one can find a representation of the semigroup associated with the process $\xi(t)$ in terms of the original path integral.
 
How this can be done was shown in \cite{Storchak_11,Storchak_2}. There, at first a similar equality was rewritten for the Green functions -- the kernels of the corresponding semigroups. Then, after introducing  the local finite covering of the manifold $\mathcal P$, the inversion of the equality was performed on charts of $\mathcal P$ that are related with the charts of the principal fiber bundle 
${\rm P}(\mathcal M, \mathcal G)$. The gluing of the local Green functions $G^{\lambda}_{mn}$ was carried out using the transition coordinate functions defined on charts of the principal fiber bundle. 
Thus, the local relations between the Green's functions, which were established by inverted equality, can be extended to global ones.
Therefore, it allows us to find the relation for the Green functions given on global manifolds, as well as for path integrals  corresponding to them.

In our case, when we are dealing with the manifold $\tilde{\cal P}$ and the principal fiber bundle 
${\rm P}(\tilde {\mathcal M}, \mathcal G)$, a similar approach leads to the  relation between the Green functions and, therefore,  to  corresponding relation between path integrals. For  Green functions, this relation is as follows:
 \begin{eqnarray}
&&G^{\lambda}_{mn}(x_b,\tilde f_b,t_b;
 x_a,\tilde f_a,t_a)=
\displaystyle\int _{\cal G}G_{\tilde{\cal P}}(p_b\theta,v_b\theta,t_b;
p_a,t_a) 
D_{nm}^\lambda (\theta )d\mu (\theta ),
\nonumber\\
&&(x,\tilde f)=\pi'(p,v).
\label{green_funk_relat}
\end{eqnarray}
The path integral which represent the Green function $G^{\lambda}_{mn}$ is written symbolically as
\begin{eqnarray}
&&G^{\lambda}_{mn}(\pi'(p_b),\pi'(v_b),t_b;
 \pi'(p_a),\pi'(v_a),t_a)=
\nonumber\\
&&{\tilde {\rm E}}_{{\xi (t_a)=\pi' (p_a,v_a)}\atop  
{\xi (t_b)=\pi' (p_b,v_b)}}
\Bigl[(\overleftarrow{\exp })_{mn}^\lambda 
(\xi(t),t_b,t_a)
\exp \{\frac 1{\mu ^2\kappa m}\int_{t_a}^{t_b}
\tilde{V}(\xi_1(u),\xi_2(u))du\}\Bigr]
\nonumber\\
&&=\int\limits_{{\xi (t_a)=\pi' (p_a,v_a)}\atop  
{\xi (t_b)=\pi' (p_b,v_b)}} d{\mu}^{\xi}
\exp \{\frac 1{\mu ^2\kappa
m}\int_{t_a}^{t_b}\tilde{V}(x(u),\tilde f(u))du\}
\nonumber\\
&&\times
\overleftarrow{\exp }%
\int_{t_a}^{t_b}\Bigl\{{\mu}^2\kappa\Bigl[\frac 12d^{\alpha \nu
}(x(u),\tilde f(u))(J_\alpha
)_{pr}^\lambda (J_\nu )_{rn}^\lambda \Bigr.\Bigr.
\nonumber\\
&&-\Bigl.\Bigl.\frac 12\frac 1{\sqrt{d\, H}}\frac \partial {\partial x^k}\left( \sqrt{d\, H}%
h^{km}\underset{\scriptscriptstyle{(\gamma)}}{{\mathscr A}_m^\nu}(x(u)) \right) (J_\nu )_{pn}^\lambda  
\nonumber\\
&&-\frac12(G^{EC}\Lambda^{\nu}_E\Lambda^{\mu}_C)\frac 1{\sqrt{d\, H}}\frac \partial {\partial \tilde f^b}\left(\sqrt{d\, H}K^b_{\mu}\right)(J_\nu )_{pn}^\lambda \Bigr]du
\nonumber\\
&&-\mu\sqrt{\kappa}\Bigl[\underset{\scriptscriptstyle{(\gamma)}}{{\mathscr A}^{\nu}_k}(x(u))\tilde X^k_{\bar m}(u)(J_\nu )_{pn}^\lambda d\tilde w^{\bar m}(u)+\tilde {\mathscr A}^{\nu}_a\tilde X^a_{\bar b}(u)(J_\nu )_{pn}^\lambda d\tilde w^{\bar b}(u)\Bigr]\Bigr\}
\label{path_int_G_mn}
\end{eqnarray}
The semigroup having this kernel acts in the space of th equivariant functions on $\tilde{\mathcal P}$:
\[
 \tilde\psi_n(pg,vg)= D^{\lambda}_{mn}(g)\tilde\psi_n(p,v).
\]
The isomorphism of these functions with the functions $\psi_n\in  \Gamma ({\tilde{\cal M}},V^{*})$
is given by
\[
 \tilde\psi_n(F(Q^{\ast}(x),e),\bar D^b_c(e)\tilde f^c)=\psi_n(x,\tilde f).
\]

We can consider the  transformation of the original path integral (\ref{orig_path_int})  leading to the integral relation between the Green functions $\mathcal G_{\tilde P}$ and $G^{\lambda}_{mn}$ which 
are the kernels of the corresponding semigroups,  as 
the reduction procedure that is performed in the path integral for the dynamical system with a symmetry. The semigroup having the Green's function $G^{\lambda}_{mn}$ as a kernel describes the "quantum" evolution of a reduced dynamical system that occurs when the initial dynamical system, in accordance with the theory of constrained systems, is reduced to a nonzero-momentum level.
In the next section  will be considered  the particular case of the path integral reduction -- reduction  to the zero-momentum level.

\section{Reduction to zero-momentum level}
 In order to get  the representation of the Green function in terms of the path integral in the case of reduction to zero-momentum level, we first   put $\lambda=0$ in  (\ref{path_int_G_mn}). 
This means that in this case we are dealing with the scalar Green functions. Now the reduced Green function describes the ``quantum'' evolution on the manifold $\tilde{\mathcal M}$ -- the orbit space of the principal fiber bundle. 
 As follows from (\ref{dif_gen_ksi}), the infinitesimal  generator of the stochastic process $\xi(t)$ of this case is the 
Laplace--Beltrami operator on $\tilde{\mathcal M}$, ${\triangle}_{\tilde{\mathcal M}}$, together with the terms that are linear in the partial derivatives with respect to $x$ and $\tilde {f}$. 

The stochastic process $\xi(t)$ is locally presented by two processes $(x^i(t),\tilde f^a(t))$: 
\[
 d\xi^i_{\rm loc}(t)=\frac12\mu^2\kappa{b^i\choose b^a}dt+\mu\sqrt{\kappa}{\tilde X^i_{\bar m}\;\; 0\choose {\tilde X^a_{\bar m}\;\; \tilde X^a_{\bar b}}} {d\tilde w^{\bar m}\choose d\tilde w^{\bar b}}.
\]
The drift coefficients $b^i$ and $b^a$ of this equation include the  terms that depend on  partial derivatives of  $d$ which is the determinant of the metric $d_{\alpha\beta}(x,\tilde f)$ given on the orbit of the principal fiber bundle:
\[
 {b^i\choose b^a}={\tilde b^i\choose \tilde b^a}+{h^{ij} \;\: \;\;\;\;\;\;\;\;\;\underset{\scriptscriptstyle{(\gamma)}}{{\mathscr A}^{\mu}_m} K^b_{\mu} h^{mi} 
\choose { \underset{\scriptscriptstyle{(\gamma)}}{{\mathscr A}^{\mu}_m} K^a_{\mu} h^{nj} \;\; G^{AB}N^a_AN^b_B+G^{ab}}}{\frac{1}{\sqrt{d}}\frac{\partial}{\partial x^j}\sqrt{d}\choose \frac{1}{\sqrt{d}}\frac{\partial}{\partial \tilde f^b}\,\sqrt{d} \; }.
\]
It is these terms lead to the additional terms of the infinitesimal  generator for the process $\xi(t)$.

One can get rid of these additional terms by using the path integral transformation known as the Girsanov transformation. In our case, this can be done as follows. 

The stochastic process $\xi(t)$  should be replaced  for a new process $\tilde\xi(t)$ with the  stochastic differential equations
\[
 d\tilde\xi^i_{\rm loc}(t)=\frac12\mu^2\kappa{\tilde b^i\choose \tilde b^a}dt+\mu\sqrt{\kappa}{\tilde X^i_{\bar m}\;\; 0\choose {\tilde X^a_{\bar m}\;\; \tilde X^a_{\bar b}}} {d\tilde w^{\bar m}\choose d\tilde w^{\bar b}}.
\]
 This transformation  changes the measure
${\mu}^{{\xi}}$ for the measure 
${\mu}^{\tilde{\xi}}$ in such a way that the Radon-Nicodim derivative of the measure ${\mu}^{\xi}$ with respect to the measure ${\mu}^{\tilde{\xi}}$ will be as follows:
\begin{eqnarray}
&&\!\!\!\!\!\!\!\!\!\frac{d{\mu}^{{\xi}}}
{d{\mu}^{\tilde{\xi}}}
({\tilde{\xi}}(t))=
\exp\int^t_{t_a}\Bigl[\mu^2\kappa<A^{-1}\frac12(b-\tilde b),dw>-\frac12(\mu^2\kappa)^2||A^{-1}\frac12(b-\tilde b)||^2 dt\Bigr], 
\nonumber\\
\label{girs}
\end{eqnarray}
where in our case we have
\[
 \mu^2\kappa<A^{-1}\frac12(b-\tilde b),dw>=\frac14\mu\sqrt{\kappa}\Bigl[(\tilde X^i_{\bar m}\,\sigma_i+\tilde X^a_{\bar m}\,\sigma_{a})d\tilde w^{\bar m}+\tilde X^a_{\bar b}\,\sigma_a\,d\tilde w^{\bar b}\Bigr]
\]
with $\sigma_i=\frac{\partial}{\partial x^i} (\ln d)$ and $\sigma_a=\frac{\partial}{\partial \tilde f^a} (\ln d)$. The second term under the integral in (\ref{girs}) is given by
\begin{equation}
 \begin{split}
&||A^{-1}\frac12(b-\tilde b)||^2  =  \\
&\frac1{16}\Bigl[h^{ij}\sigma_i\sigma_j+2h^{kj}\underset{\scriptscriptstyle{(\gamma)}}{{\mathscr A}^{\mu}_k} K^a_{\mu} \sigma_a\sigma_j+\Bigl((\gamma^{\alpha\beta}+h^{kl}\underset{\scriptscriptstyle{(\gamma)}}{{\mathscr A}^{\alpha}_k}  \underset{\scriptscriptstyle{(\gamma)}}{{\mathscr A}^{\beta}_l}) K^a_{\alpha}K^b_{\beta}+G^{ab}\Bigr)\sigma_a\sigma_b\Bigr] 
\end{split}
\end{equation}

The terms with the differentials $\tilde w$  in (\ref{girs}) can be transforme with the It\^{o} identity. This identity is obtained by differentiation $\exp (\sigma(x(t),\tilde f(t))$ by It\^{o} formula. As a result we get 
\begin{eqnarray}
&& \exp\int^{t_b}_{t_a}\!\!4<A^{-1}\frac12(b-\tilde b),dw(t)>=\left(\frac{exp(\sigma(x(t_b),\tilde f(t_b))}{exp(\sigma(x(t_a),\tilde f(t_a))}\right)\nonumber\\
&&\times\exp\Bigl\{-\int^{t_b}_{t_a}\Bigl[\frac12\sigma_i\tilde b^i+\frac12\sigma_{ij}h^{ij}+\frac12\sigma_a\tilde b^a+\sigma_{ia}h^{in}\underset{\scriptscriptstyle{(\gamma)}}{{\mathscr A}^{\mu}_n} K^a_{\mu}\Bigr.
\nonumber\\
&&\Bigl.+\frac12\sigma_{ab}\Bigl((\gamma^{\alpha\beta}
+h^{kl}\underset{\scriptscriptstyle{(\gamma)}}{{\mathscr A}^{\alpha}_k}\underset{\scriptscriptstyle{(\gamma)}}{{\mathscr A}^{\beta}_l}) K^a_{\alpha}K^b_{\beta}+G^{ab}\Bigr)\Bigr]dt\Bigr\}
\label{ident_ito}
\end{eqnarray}
In this expression  $\tilde b^i$ and ${\tilde b^a}$ are given by
\[
 \tilde b^i=\frac1{\sqrt{H}}\frac{\partial}{\partial x^j}\Bigl(\sqrt{H}h^{ij}\Bigr)+\underset{\scriptscriptstyle{(\gamma)}}{{\mathscr A}^{\mu}_n}h^{ni}\frac1{\sqrt{H}}\frac{\partial}{\partial \tilde f^b}\Bigl(\sqrt{H}K^b_{\mu}\Bigr),
\]
and
\begin{eqnarray*}
 \tilde b^a&=&\frac1{\sqrt{H}}\frac{\partial}{\partial x^j}\Bigl(\sqrt{H}h^{mj}\underset{\scriptscriptstyle{(\gamma)}}{{\mathscr A}^{\mu}_m}\Bigr)K^a_{\mu}+(G^{ab}+G^{AB}N^a_AN^b_B)\frac1{\sqrt{H}}\frac{\partial}{\partial \tilde f^b}\Bigl(\sqrt{H}\Bigr)
\\
&&+\frac{\partial}{\partial \tilde f^b}\Bigl(G^{AB}N^a_AN^b_B\Bigr).
\end{eqnarray*}

The result of the Girsanov  transformation is the following:
 \begin{eqnarray}
\!\!\!\!\!\!\!\!\!\frac{d{\mu}^{{\xi}}}
{d{\mu}^{\tilde{\xi}}}
({\tilde{\xi}}(t))&=&\left(\frac{exp(\sigma(x(t_b),\tilde f(t_b))}{exp(\sigma(x(t_a),\tilde f(t_a))}\right)^{1/4}
\nonumber\\
&&\times\exp\Bigl\{-\frac18\mu^2\kappa\int^{t_b}_{t_a}\bigl(\triangle_{\tilde{\cal M}}\sigma +\frac14<\partial\sigma,\partial \sigma>_{\tilde{\cal M}}\bigr)du\Bigr\}
\label{girs_result}
\end{eqnarray}
where by $<\partial\sigma,\partial \sigma>_{\tilde{\cal M}}$ we denote the following quadratic form obtained with the inverse metric  on $\tilde{\cal M}$:
\[
 \Bigl[h^{ij}\sigma_i\sigma_j+2h^{kj}\underset{\scriptscriptstyle{(\gamma)}}{{\mathscr A}^{\mu}_k} K^a_{\mu} \sigma_a\sigma_j+\Bigl((\gamma^{\alpha\beta}+h^{kl}\underset{\scriptscriptstyle{(\gamma)}}{{\mathscr A}^{\alpha}_k}  \underset{\scriptscriptstyle{(\gamma)}}{{\mathscr A}^{\beta}_l}) K^a_{\alpha}K^b_{\beta}+G^{ab}\Bigr)\sigma_a\sigma_b\Bigr].
\]
We note the reduction Jacobian is similar in form to  that which was  obtained earlier in \cite{Storchak_1,Storchak_11} for the case of the mechanical system related with the  principal fiber bundle $\rm P(\cal M,\cal G)$. The Jacobian is also coinsides by its structure with the quantum potential obtained in \cite{Lott} for the same principal bundle.

Thus, in the case of the reduction to the zero-momentum level we get the following integral relation:
\begin{equation}
d_b^{-1/4}d_a^{-1/4}
G_{\tilde{M}}(x_b,\tilde f_b, t_b;x_a,\tilde f_a,t_a)=\int_{\mathcal G}{G}_{\tilde{\mathcal P}}(p_b\theta,v_b\theta,t_b;p_a,v_a,t_a)
d\mu (\theta ),
\nonumber\\
\end{equation}
where we use the following notation: $d_b=d(x_b,\tilde f_b)$, $d_a=d(x_a,\tilde f_a)$.

The semigroup which is determined by the Geen function $G_{\tilde{M}}$ acts in the Hilbert space with the scalar product $(\psi_1,\psi_2)=\int \psi_1(x,\tilde f),\psi_2(x,\tilde f)dv_{\tilde{\mathcal M}}.$

The Green function $G_{\tilde{M}}$ is presented by the path integral as follows: 
\begin{eqnarray*}
&&G_{\tilde{\mathcal M}}(x_b,\tilde f_b, t_b;x_a,\tilde f_a,t_a)
\nonumber\\
&&\;\;\;\;\;\;\;\;\;\;\;\;\;=\int_{ 
{\tilde{\xi}(t_a)=(x_a,\tilde f_a)}\atop
{\tilde{\xi}(t_b)=(x_b,\tilde f_b)}}
d\mu ^{\tilde{\xi}}\exp 
\left\{\frac 1{\mu
^2\kappa m}\int_{t_a}^{t_b}
V(\tilde{\xi}_1(u),\tilde{\xi}_2(u))du\right\}
\nonumber\\
&&\;\;\;\;\;\;\;\;\;\;\;\;\;\times\exp\Bigl\{-\frac18\mu^2\kappa\int^{t_b}_{t_a}\bigl(\triangle_{\tilde{\cal M}}\sigma +\frac14<\partial\sigma,\partial \sigma>_{\tilde{\cal M}}\bigr)du\Bigr\},
\nonumber\\
&&\;\;\;\;\;\;\;\;\;\;\;\;\;(x,\tilde f)=\pi'(p,v),  \;\;\sigma =\sigma(\tilde{\xi}_1(u),\tilde{\xi}_2(u)).
\end{eqnarray*}

The transition to the Schr\"odinger equation is performed from the forward Kolmogorov equation. 
The Green function $G_{\tilde{\mathcal M}}$ satisfies this equation with respect to  the  variables $(x_b,\tilde f_b,t_b)$. The operator of
the forward Kolmogorov equation is
\[
\hat{H}_{\kappa}=
\frac{\hbar \kappa}{2m}\triangle _{\tilde{M}}-\frac{\hbar \kappa}{8m}\Bigl[\triangle_{\tilde{\cal M}}\sigma +\frac14<\partial\sigma,\partial \sigma>_{\tilde{\cal M}}\Bigr]+\frac{1}{\hbar \kappa}V.
\]

At $\kappa =i$ this forward Kolmogorov equation becomes
the Schr\"odinger equation with the Hamilton operator 
$\hat H=-\frac{\hbar}{\kappa}{\hat H}_{\kappa}\bigl|_{\kappa =i}$.

\section{Conclusion}
In the article it was shown the reduction procedure in the path integrals can also  be applied to the principal fiber bundles having the  more complex structure. The resulting reduction Jacobian for the case of the reduction to the zero-momentum level has the same form as the Jacobian for the reduction associated with the standard principal bundle whose base space is an ordinary manifold. This means that the resulting Jacobian has the same  geometric representation. 

We also note that using the same approach, it  is possible to  realize the path integral reduction  in case of the nonzero-momentum level. In this case, applying the Girsanov transformation, the transformation of the multiplicative functional should be taken into account.

In conclusion, we note the problems concerning with the transition to global expressions from local ones.
The assumption that by the local expressions one can  able to restore the global expressions  can only be true in some cases. First of all, this is the case when the principal fiber bundle $\rm P(\mathcal M, \mathcal G)$ is trivial. It takes place, for example, when the local submanifolds $\{\chi ^{\alpha}=0\}$ form the global submanifold of the  manifold $\mathcal P$. Note that  
 ${\rm P}(\mathcal P\times_{\mathcal G} V, \mathcal G)$ will be also trivial.

As a consequence, we come to  a local isomorphism of the trivial principal  fiber bundle $\rm P(\mathcal M,\mathcal G)$ and the trivial principal  bundle 
$\pi_{\Sigma}:\Sigma \times \mathcal G \to \Sigma$ 
\cite{Mitter-Viallet,Huffel-Kelnhofer}. Therefore,  the charts of  the total space $\mathcal P$ are expressed through the charts of the global submanifold $\Sigma$. 
And  constrained global variables, defined on $\Sigma$, can be used as the coordinate functions of these charts.
It folows that  in this case, for the trivial principal fiber bundle ${\rm P}(\mathcal P\times_{\mathcal G} V, \mathcal G)$, we  have a bundle  isomorphism $\tilde \varphi: \tilde \Sigma\times \mathcal G\to \mathcal P\times V$ which enables us to define the charts with adapted coordinates on this bundle.

 The second possibility arises when having non-trivial principal bundle,  we restrict our consideration to the domain defined by the submanifold  $\Sigma$.  This is the case of the gauge theories, where the gauges are "unresolved", i.e., the cannot be paramerically given. And considerstion forsed to be restricted by the Gribov horizon. The principal bundle coordinates are introduced as in \cite{Mitter-Viallet}.

Finally, it can be assumed  the case when there is a sufficient number of consistent between themselves surfaces with the necessary properties. Then these surfaces plays the role of the sections in the principal fiber bundle. 
Then the principal bundle coordinates are determined in a standard way.
 
Despite its relevance, mainly for  gauge theories, a full understanding of these issues has not yet been achieved. Therefore, further research is needed to further study these problems.

\end{document}